\documentstyle[12pt,psfig]{article}
\begin{document}
\title{Electron Propagation in the Field of Colliding Nuclei at
Ultrarelativistic Energies}
\author{U.~Eichmann${}^a$, J.~Reinhardt${}^a$, S.~Schramm${}^b$,
W.~Greiner${}^a$}
\date{${}^a$Institut f\"ur Theoretische Physik,\\
Johann Wolfgang Goethe-Universit\"at, Frankfurt am Main, Germany\\
${}^b$Gesellschaft f\"ur Schwerionenforschung mbH, Darmstadt, Germany }
\maketitle
\begin{abstract}
We calculate the asymptotic high-energy amplitude
for electrons scattering at one ion as well as at two colliding ions,
respectively, by means of perturbation theory. We show 
that the interaction with one ion {\it eikonalizes} and that the interaction
with two ions {\it causally decouples}. We are able to put previous results
on perturbative grounds and propose further applications for the obtained
rules for interactions on the light cone. The formalism will be of use for
the calculation of Coulomb corrections to electron-positron pair creation in
heavy ion collisions. Finally we discuss the results and 
inherent dangers of the employed approximations. 
\end{abstract}

\section{Introduction}
At ultrarelativistic energies, the theoretical treatment of scattering
processes is extremely facilitated. On the one hand, the relevant equations
themselves simplify, when terms of order ${\cal O}(1/\gamma^2)$ become
negligible, on the other hand, the interactions simplify due to causality. 
In that way, high energy scattering becomes analytically
accessible. 

Eikonal approximations or optical models usually are formulated for 
the scattering of a highly energetic particle at a slow or even 
static center
\cite{Abarbanel}\cite{Torgerson}. We present a
simple transformation of the covariant derivatives 
that is used to easily solve the opposite case. 
The transformation of the equations of motion for particles
scattered by fast moving charge centers immediately generates the scattered wave
describing the particle. Our results coincide with previous calculations
performed in this reference frame
\cite{Jackiw}\cite{tHooft}. 

The summation of ladder graphs is shown to {\it eikonalize} as well
\cite{Chang-Ma}.
This was elegantly derived within 
the method of kinematically decoupling the components of the 
scattering process, and
Lorentz transforming into the respective rest frames \cite{Chang-Fishbane} 
which inherently contains the advantages of a fast external potential. 

Following a different approach  we will exploit the same advantages. 
We perform a perturbative approach and directly approximate the 
external potential by its
asymptotic high-energy limit which amounts to saying, that the longitudinal
components of the exchanged photons can be discarded.

In doing so, one can directly rederive the amplitude for the
scattering at one center and even put the recent result of Segev and Wells 
\cite{Segev} for the scattering amplitude for an electron moving in the
field of two ultrarelativistic colliding ions on perturbative grounds. 
Moreover, one is allowed to go beyond their calculations and
is provided deeper insight. 

The derivations in this paper are formulated for electron scattering, but
they can be immediately extended to cover the physically more relevant
process of electron-positron pair production. 
The search for exact analytic expressions describing 
electron-positron pair production in heavy-ion collisions 
is motivated by the question whether 
Coulomb effects only play an inferior role at high energies. Such a
conclusion might be drawn  
from a comparison between second-order 
perturbation theory results \cite{Bottcher} and calculations employing
Furry-Sommerfeld-Maue
wave functions \cite{Ionescu}. 
It should be mentioned, however, that the Coulomb distortions
considered in these calculations 
only account for one ion, whereas the second ion enters as a perturbation.  

\section{Scattering of an electron off a fast moving source}
\subsection{Transformation of the Dirac equation}
\label{1iontrafo}
We are searching for the asymptotic scattering solution of a Dirac particle 
from a 
fast moving Coulomb potential in the limit of very large collision energy. 
In the Lorentz gauge the Li$\acute{\rm e}$nard-Wiechert potentials for a point 
charge moving moving with uniform velocity $\beta$ in $+z$ direction read 
\begin{eqnarray}
\label{lwpot}
A_0&=&-\frac{Z\alpha \gamma}{\sqrt{\gamma^2(z-\beta t)^2 +
\vec{x}_\perp^2}}\\
\label{a0a3}
A_3&=&\beta A_0
\end{eqnarray}
The equation of motion for the scattered particle becomes
\begin{equation}
\label{dg}
\left[ \hat{\gamma_0}(i\partial_t -A_0)+\hat{\gamma_3}(i\partial_z +A_3)+
\hat{\vec{\gamma}}_\perp \cdot  i\vec{\nabla}_\perp - 
m\right] \psi =0
\end{equation}
We set $c=\hbar=1$. The charge $e$ of the electron was absorbed into the
definition of the potential. 
We make use of the external field approximation, i.e. 
we assume that the source is not
influenced by the scattered particle and moves on a straight line. 
This treatment will be justified if
the mass of the source particle is very large. 
To simplify the Dirac equation (\ref{dg}) we use the operator identity
\cite{Miura}
\begin{equation}
(i\partial_x \mp
i\partial_x\ln \phi)^n=\phi^{\pm 1}(i\partial_x)^n\phi^{\mp 1}
\label{kovabl}
\end{equation}
to rewrite the covariant derivatives.
We must introduce two fields $\phi'$ and $\phi$ for the
space and the time component of the vector-potential $A^\mu$  
\begin{eqnarray}
A_0&=&i\partial_t \ln \phi\nonumber\\
A_3&=&i\partial_z \ln \phi'
\end{eqnarray}
The field $\phi$ is determined to be
\begin{equation}
\label{phidef}
\phi=e^{-i\int_{-\infty}^t dt' A_0}
\end{equation}
The thus transformed
Dirac equation  reads (see Appendix \ref{adiractrans})
\begin{equation}
\label{dtrans}
\left[\hat{\gamma}_0i\partial_t +\hat{\gamma}_3 
(i\partial_z-\frac{1}{\beta^2\gamma^2}A_3)+
\hat{\vec{\gamma}}_\perp \cdot (i\vec{\nabla}_\perp + 
i{\rm grad}_\perp
\ln \phi )- m\right]\tilde{\psi}=0
\end{equation}
where $\tilde{\psi}=\psi/\phi = e^{i\int_{-\infty}^t dt' A_0}\psi$. The
operator identity, together with the field $\phi$ defined in (\ref{phidef})
has led to the elimination of the 
scalar part
of the vector potential, i.e. to the temporal gauge, $A'_0=0$. 
For very large $\gamma$ one nearly has a purely transverse vector potential
$\vec{A'}_\perp = i{\rm grad}_\perp
\ln \phi$ 
which is the negative time integral of the transverse electric field. 
From classical 
electrodynamics one knows, that the time 
integral of the transverse electric field is given by 
\begin{equation}
\label{etransint}
\int_{-\infty}^{\infty}\vec{E}_\perp=-2Z\alpha\vec{x}_\perp /
(\beta x_\perp^2)
\end{equation} 
This implies that 
\begin{equation}                             
\int_{-\infty}^\infty dt' A_0 = +\frac{1}{\beta}
Z\alpha \ln x_\perp^2 + {\cal C}
\end{equation}
which reproduces (\ref{etransint}) if the transverse gradient operator is
applied. 
${\cal C}$ is an infinite quantity which expresses the divergence of the
phases in Coulomb scattering. 
Furthermore using (\ref{phidef}) and (\ref{lwpot}) 
it is easy to show, that the transverse vector potential exhibits 
a Heaviside step function dependence 
$\sim \theta (t -z)\vec{x}_\perp/x^2_\perp$ in the limit 
of very large $\gamma$. Now, since $t$ and $\gamma$ enter symmetrically in the 
integral, the limit $\gamma \to \infty$ corresponds to sending the upper bound 
of the integral to infinity. Therefore, all of the above is applicable and we 
find 
\begin{equation}                                                      
\label{potlimc}
\lim_{\gamma \to \infty} A_0 = +\delta (z-t) Z \alpha \ln x_\perp^2 + {\cal
C}'
\end{equation}
The Coulomb phase ${\cal C}'$ in general will depend on $z$ and $t$. It 
can be removed by a gauge transformation, 
as is easily seen
\begin{equation}
\label{cgtrafo}
\tilde{\psi}'=e^{-i\int_{-\infty}^{t}dt'{\cal C}'}\tilde{\psi}=
e^{+iZ\alpha \theta (t-z) 
\ln x_\perp^2}\psi
\end{equation}
This gauge transformation was first applied in \cite{Aichelburg}. 
The removal of the Coulomb phase yields a short range potential allowing for
asymptotic plane wave solutions (see Appendix \ref{apotlim}). 

For $t\neq z$ the $t$ and $z$ dependence in both the transverse vector 
potential and the transformed spinor $\tilde{\psi}$ vanish in the limit
$\gamma\to \infty$. By inverse transformation we find 
that $\psi$ solves a 
free Dirac equation on either side of the light front $t=z$ 
and can only differ by a phase.  

The transformed wave function $\tilde{\psi}$ has the advantage of being 
continuous on the surface  defined by 
$t=z$. In contrast, the wave function $\psi$ exhibits a discontinuous
behaviour at the light front. 
There is a jump in that component of $\psi$ which
couples to 
$\hat{\gamma}_-
=\hat{\gamma}_0-\hat{\gamma}_3$, the matrix structure of the interaction in
the limit $\gamma \to \infty$. Using this 
property one directly finds for $\hat{\gamma}_-\psi$ at the discontinuity 
\begin{equation}
\label{psiout}
\hat{\gamma}_-\psi(t-z=0^+)=e^{-iZ\alpha \ln x_\perp ^2 }\hat{\gamma}_-\psi
(t-z=0^-)
\end{equation}
where we ignored the irrelevant quantity ${\cal C}$.\footnote{The effect of
the potential (\ref{potlimc}) also can be described within the
Aichelburg-Sexl metric. Two field-free regions of space-time meet at $z=t$,
such that (the superscripts $<$ and $>$ denote $t>z$ and $t<z$,
respectively)
\protect\begin{eqnarray}
\label{xtrgr}x_\perp^>&=&x_\perp^<\\
\label{zgr}z^>&=& z^<-Z\alpha \ln x_\perp^2\\
\label{tgr}t^>&=& t^<-Z\alpha \ln x_\perp^2
\end{eqnarray}
The result (\ref{psiout}) is then easily obtained by simply substituting 
(\ref{xtrgr})-(\ref{tgr}) into the plane wave at $t>z$.} The complement
$\hat{\gamma}_+\psi$ of
these spinor components, where $\hat{\gamma}_+ = 2\hat{\gamma}_0
-\hat{\gamma}_-= \hat{\gamma}_0+\hat{\gamma}_+$  
is continuous
at $t=z$. Both parts of the spinor are coupled via the free Dirac equation
on either side of the discontinuity. 

By application of the LSZ-reduction formula 
one finds in general, that at very large scattering energies the $S$ 
matrix is determined by $\phi$, in which we recognize the well 
known eikonal form \cite{Fried}. 
Because of the identity (\ref{kovabl}) 
this result holds independently of the power of the momentum 
entering in the respective wave equation. 
For that reason the expressions for the 
$S$ matrices for e.g. spinor or scalar particles 
only differ by an overall factor. 

We first consider the properties of the previous result. 
The LHS of 
(\ref{psiout}) can be expanded in plane waves. Since we consider scattering
at the negative light front, we must substitute $d^3x \to dx_+d^2x_\perp$
\cite{Meggiolaro} and accordingly $d^3p \to dp_-d^2p_\perp$. 
The expansion coefficient corresponds to the $S$ matrix in momentum space,
which is easily found to be 
\begin{eqnarray}
\label{smom}
S(p',p)&=&2\pi\delta
(p'_--p_-)
\Bigg[\left(\frac{4}{(\vec{p'}_\perp-\vec{p}_\perp)^2}
\right)^{1-iZ\alpha}\Gamma^2(1-iZ\alpha)\sin(\pi
iZ\alpha)\nonumber\\
&&+(2\pi)^2\delta(\vec{p'}_\perp-\vec{p}_\perp)\Bigg]
\overline{u}(p')\hat{\gamma}_-u(p)
\end{eqnarray}
Here $u$ denotes the electron unit spinor. 
If the trajectory of the ion is shifted by the impact parameter $\vec{b}$,
this result is simply multiplied by the factor 
$e^{i(\vec{p'}_\perp-\vec{p}_\perp)\cdot \vec{b}}$.
$p$ and $p'$ are the incoming and outgoing momenta. 
We note that 
the negative light cone momentum $p_-=p_0-p_3$ is conserved in the
scattering. The positive
light cone momentum is fixed by the mass shell condition. 
The first term in the square brackets in  
(\ref{smom}) corresponds to the $T$ matrix. 
Eq. (\ref{smom}) represents a well known result which was 
previously derived in e.g. 
\cite{Abarbanel,tHooft,Jackiw,Segev}. 

\subsection{Perturbative Approach}
\label{pertapp}
\begin{figure}[htbp]
\centerline{\psfig{figure=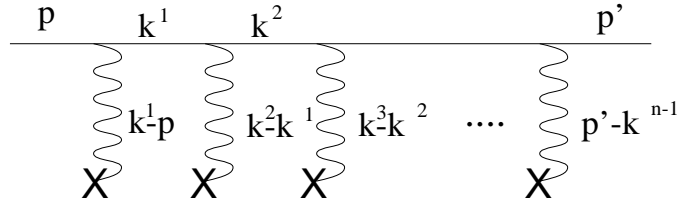,angle=270,width=9cm}}
\caption{\label{fig01}Scattering of an electron at an external potential. }
\end{figure}

In this section we want to derive the eikonal form of the $T$ matrix via 
perturbation theory. 
Several approximations are necessary to obtain the
eikonal form, namely the neglect of the longitudinal components of the
photon momentum, the conservation of the 
photon light cone momentum, as well as the simplification of the
matrix structure of the interaction \cite{Chang-Ma}. The calculation shows that these 
approximations are the counterparts of the
requirement of a 
vanishing longitudinal vector potential and the step function dependence of
the transverse vector potential. 
Having this in mind we directly use the asymptotic high-energy 
expression of the potential. 
We then evaluate the terms of the perturbation series for the external-field
scattering problem depicted by the Feynman graphs of Fig. \ref{fig01}.

The potential entering into our calculations is of the form
\begin{equation}
\label{pot}
V_0(x)=V_3(x)=\delta (z-t)V_\perp(\vec{x}_\perp)
\end{equation}
In the following calculations it will not be necessary to specify the
explicit form of $V_\perp(\vec{x}_\perp)$. Problems related to the
logarithmic potential obtained in the last section will be discussed in
section \ref{discussion}.
We use the light-cone variables 
\begin{equation}
\left(\begin{array}{c}p_-\\ p_+\end{array}\right)=\left(\begin{array}{rr} 
1&-1\\ 1&1\end{array}\right)\left(\begin{array}{c}p_0\\ p_3\end{array}\right)
\end{equation}
The Feynman propagator describing the internal electron lines reads
\begin{equation}
S_F(p)=\frac{1}{\hat{\gamma}_0p_0 -\hat{\vec{\gamma}}\cdot\vec{p}-
m+i\epsilon}=
\frac{\frac{1}{2}(\hat{\gamma}_-p_++\hat{\gamma}_+p_-)- 
\hat{\vec{\gamma}}_\perp 
\cdot \vec{p}_\perp +m}{p_+p_--p_\perp^2-m^2+i\epsilon}
\end{equation}
The following products of gamma matrices in the light-cone representation
are needed:
\begin{equation}
\label{gamprod}
\begin{array}{ll}
\displaystyle
\hat{\gamma}_\pm \hat{\gamma}_\mp \hat{\gamma}_\pm &
\displaystyle 
=4\hat{\gamma}_\pm\\
\displaystyle
\hat{\gamma}_\pm \hat{\vec{\gamma}}_\perp \hat{\gamma}_\mp&\displaystyle 
 = 2\hat{\vec{\alpha}}_\perp \hat{\gamma}_\mp\\
\displaystyle 
\hat{\gamma}_\pm \hat{\vec{\gamma}}_\perp \hat{\gamma}_\pm
&\displaystyle  =0\\
\displaystyle 
\hat{\gamma}_\pm \hat{\gamma}_\mp&\displaystyle 
=2\hat{\gamma}_0\hat{\gamma}_\mp\\
\displaystyle
\hat{\gamma}_\pm \hat{\gamma}_\pm&\displaystyle 
=0\\
\end{array}
\end{equation}
The amplitude for electron scattering in first order perturbation theory is
\begin{equation}
\label{1order}
A^{(1)}_{p'p}=(2\pi) (-i) \delta (p'_--p_-)
F_{p'p}(V_\perp)\overline{u}(p')\hat{\gamma}_-u(p)
\end{equation}
$F_{p'p}(\;\;)$ denotes the Fourier transform with respect to the transverse
coordinates taken at the momentum $(\vec{p'}_\perp-\vec{p}_\perp)$
\begin{equation}
F_{p'p}(V_\perp)=\int d^2 x_\perp e^{-i\vec{x}_\perp \cdot 
(\vec{p'}_\perp-\vec{p}_\perp)}V_\perp (\vec{x}_\perp)
\end{equation}
In second order the amplitude reads
\begin{eqnarray}
A^{(2)}_{p'p}&=&\int \frac{dk_+dk_-d^2k_\perp}{(2\pi)^4} (2\pi)^{2} 
(-i)^2 i \delta (k_--p_-)\delta
(p'_--k_-)\frac{k_-}{k_-k_+-k_\perp^2 -m^2
+i\epsilon}\nonumber \\  
\label{k+int}
&&F_{kp}(V_\perp)F_{p'k}(V_\perp)\overline{u}(p')\hat{\gamma}_-u(p)\\
 &=&(2\pi) (-i)^2 \delta (p'_--p_-)
\frac{1}{2}F_{p'p}(V_\perp^2)\overline{u}(p')\hat{\gamma}_-u(p)
\nonumber
\end{eqnarray}
The $k_+$ integral in (\ref{k+int}) drops out using the symbolic
substitution 
\begin{equation}
\label{Feynsubst}
1/(x+i\epsilon)\to P(1/x)-i\pi\delta (x)
\end{equation}
since the principal value integral $P$ vanishes. 
It is interesting to note that the simple structure of the results
(\ref{1order}) and (\ref{k+int}) is retained if one goes to higher orders of
perturbation theory. 
The $n^{th}$ order amplitude factorizes into $n-1$ integrals of the form
(\ref{k+int}) which leads to  
\begin{equation}
\label{norder}
A^{(n)}_{p'p}=(2\pi) (-i)^n \delta (p'_--p_-)
\frac{1}{n!}F_{p'p}(V_\perp^n)\overline{u}(p')
\hat{\gamma}_-u(p)
\end{equation}

This result is obtained by symmetrizing the $n-1$ 
integrals over the positive light cone momenta in (\ref{Feynsubst}) yielding 
the expression
$(-i2\pi)^{n-1}/n!\prod_i \delta (k_+^i)$ \cite{Chang-Ma}. This corresponds to 
reconsidering the
different time orderings and finally dividing by $n!$ to prevent double
counting. This symmetrization procedure directly shows that the
principal value terms in (\ref{Feynsubst}) do not contribute.

Clearly, with (\ref{norder}) 
the perturbation series can be summed up to yield the result 
\begin{equation}
\label{tmatr1}
A_{p'p}=2\pi\delta (p'_--p_-){\cal
T}(\vec{p'}_\perp-\vec{p}_\perp)\overline{u}(p')\hat{\gamma}_-u(p)
\end{equation}
Here we defined the momentum transfer function 
\begin{equation}
{\cal
T}(\vec{p'}_\perp-\vec{p}_\perp)=F_{p'p}(e^{-iV_\perp(\vec{x}_\perp)}-1)
\end{equation}
with
\begin{equation}
V_\perp(\vec{x}_\perp)=\int_{-\infty}^{+\infty} dt V_0(x)
\end{equation}
This result reproduces the eikonal form.

\section{Solution in the field of several ions}
\subsection{The case of two colliding ions}
\begin{figure}[htbp]
\centerline{\psfig{figure=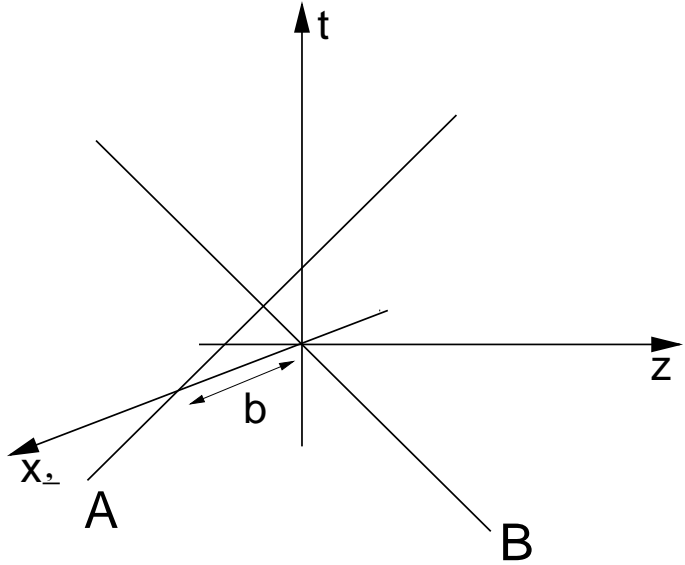,angle=270,width=9cm}}
\caption{\label{fig02}Geometry given by two lightlike ions colliding 
with an impact
parameter $\vec{b}$. The $x_\perp$-axis symbolically denotes the transverse
plane. The $\vec{b}$ 
dependence of (\protect\ref{pot2ion}) is absorbed in the definitions of
$V^{A,B}_\perp(\vec{x}_\perp)$. }
\end{figure}

In the c.m. frame, the field of two ultrarelativistic 
colliding ions $A$ and $B$, cf. Figure \ref{fig02}, reads 
\begin{equation}
\label{pot2ion}
V_{0/3}(x)=\delta (z-t) V^A_\perp(\vec{x}_\perp)
\pm \delta (z+t)V^B_\perp(\vec{x}_\perp)
\end{equation}

The identity (\ref{kovabl}) can also be applied to potentials
given by a superposition as is easily verified
\begin{equation}
\label{mtrafomi}
(i\partial_x \mp i\sum \partial_x\ln \phi_i)^n=\left(\prod \phi_i\right)^
{\pm 1}
(i\partial_x)^n
\left(\prod \phi_i\right)^{\mp 1}
\end{equation}
Since in the case of (\ref{pot2ion}) we have two discontinuities, the
asymptotic solution is not obtained as easily as in section \ref{1iontrafo}.
The explicit calculation is shown in Appendix \ref{asolu2ion}.

It is found, that the two ions couple to distinct components of the electron
spinor. 
We show in this section, how this behaviour follows from perturbation theory
and how it can be interpreted consistently. 

\begin{figure}[htbp]
\centerline{\psfig{figure=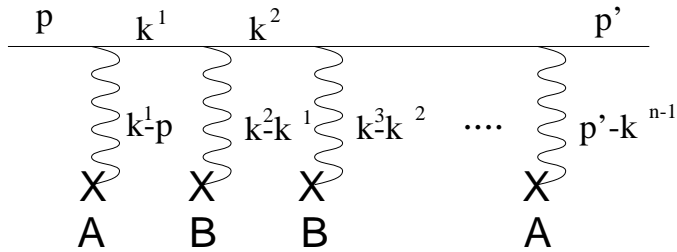,angle=270,width=9cm}}
\caption{\label{fig03}Same as Fig. \protect\ref{fig01} but 
considering two ions $A$ and
$B$ as external sources. This diagram does not contribute in the high-energy
limit.}
\end{figure}

\begin{figure}[htbp]
\centerline{\psfig{figure=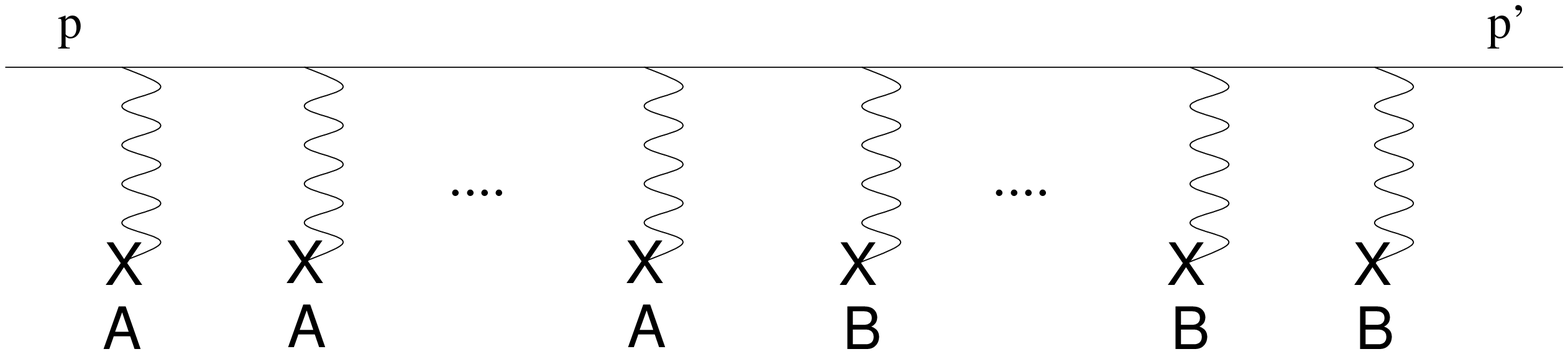,height=3cm}}
\caption{\label{fig04}Class of Feynman graphs contributing in the
high-energy limit.}
\end{figure}

We have to consider several new diagrams describing the alternate
interaction of the electron with both ions. 
Using (\ref{gamprod}) we find, that the contribution to the $T$ matrix 
of an arbitrary number of interactions with one ion that are sandwiched
between interactions with the other ion (see Fig. \ref{fig03})
, vanishes. The reason is, that we
end up with an integral of the form
\begin{equation}
A\sim\int \frac{dk_\pm}{(k_\pm p_\mp -k_\perp^2 -m^2 +i\epsilon) 
(k_\pm p'_\mp -{k'}^2_\perp -m^2 +i\epsilon)}=0
\end{equation}
The vanishing of this integral is immediately seen from Cauchy's theorem
since 
the contour can be closed in the upper half plane, where the integrand
is analytic.

In the ultrarelativistic limit the electron will therefore 
interact with the ions
separately, see Figure \ref{fig04}. 
The separate interactions of the electron with the two ions $A$ and $B$ 
are linked in the
following way 
\begin{eqnarray}
A^{tot}_{p'p}&=&\int \frac{d^2k_\perp}{(2\pi)^2}
{\cal T}_A(-\vec{p}_\perp+\vec{k}_\perp){\cal T}_B(\vec{p'}_\perp -
\vec{k}_\perp)\nonumber \\
&&\overline{u}(p')\frac{-\hat{\vec{\alpha}}_\perp 
\cdot \vec{k}_\perp + 
\gamma_0
m}{p'_+ p_- - {k}_\perp^2 -m^2 +i\epsilon} 
\hat{\gamma}_+u(p)
\nonumber\\
&&+\int \frac{d^2k_\perp}{(2\pi)^2}
{\cal T}_A(-\vec{p}_\perp+\vec{k}_\perp){\cal T}_B(\vec{p'}_\perp -
\vec{k}_\perp)\nonumber \\
\label{Ttot}
&&\overline{u}(p')\frac{-\hat{\vec{\alpha}}_\perp 
\cdot (\vec{p}_\perp +\vec{p'}_\perp -\vec{k}_\perp) +
\gamma_0
m}{p'_- p_+ - 
(\vec{p}_\perp +\vec{p'}_\perp -{k}_\perp)^2 -m^2 +i\epsilon} 
\hat{\gamma}_-u(p)
\end{eqnarray}
Here we have already added both possible time orderings. 
${\cal T}_A$ and 
${\cal T}_B$ are
the momentum transfer 
functions ${\cal T}$, defined in (\ref{tmatr1}) 
for the interactions with ion $A$ and $B$, respectively. 
This result is equivalently obtained by using the discontinuous behaviour at
the light fronts (see Appendix \ref{asolu2ion}) and corresponds to 
the result of Segev and Wells \cite{Segev}. 

To understand the decoupling property, one has to consider the matrix
structure of the potential. To this end we write down the Dirac equation in
the following form 
\begin{equation}
\left[ i\partial_t+
\hat{\vec{\alpha}} \cdot i \vec{\nabla} - \hat{\gamma}_0m
-(1\pm \beta\hat{\alpha_z})A_0 \right] \psi =0
\end{equation}
where the sign depends upon the direction of motion and  $A_0$ is given by 
(\ref{lwpot}).
In the limit $\beta\to 1$ the operators $1/2(1\pm \beta\hat{\alpha_z})$ become
orthogonal projection operators \cite{Segev}.
The
action of these operators can be understood if one recalls the standard form of Lorentz
transformations \cite{Itzykson} in spinor space
\begin{equation}
\label{ltrafo}
\psi'(x')=e^{-(i/4)\sigma_{\alpha \beta}\omega^{\alpha\beta}}\psi (x)
\end{equation}
Here $\sigma_{\alpha \beta}=i/2[\gamma_\alpha,\gamma_\beta]$ and
the exponent represents the product of 
the rapidity vector $\vec{\omega}$ times the generators of the
Lorentz transformation. For a boost in $+z$ direction (\ref{ltrafo})
simplifies to 
\begin{eqnarray}
\psi'(x')&=&e^{-\frac{\omega}{2}\hat{\alpha}_z}\psi (x)\nonumber \\
 &=&{\rm cosh}\left(\frac{\omega}{2}\right)\left(1-{\rm
tanh}\left(\frac{\omega}{2}\right)\hat{\alpha}_z\right)\psi (x)
\end{eqnarray}
Therefore, (see (\ref{lwpot})), a Lorentz-transformed vector acting in
spinor space
\[
(1\pm \beta\hat{\alpha_z})A_0=\gamma(1\pm
\beta\hat{\alpha_z})\frac{-Z\alpha}{r'}
\]
can directly be obtained by a Lorentz transformation (\ref{ltrafo})
accounting for the vectorial nature of the transformed object with a factor
2 in the exponent.
The operators $(1\pm\hat{\alpha}_z)$ are 
$1/\gamma$ times a Lorentz transformation with effectively infinite
rapidity. These operators project the Dirac spinors onto causally
disconnected subspaces of the Hilbert space. Therefore it is simply causally
impossible for the Dirac spinor to communicate alternately with both ions.

Therefore, even the exact expressions for the interaction of an electron
with two colliding ultrarelativistic ions maintains the structure of the
two-photon graph. We can interpret (\ref{Ttot}) as the interaction of an
electron in lowest order with a ''dressed`` potential of the form
\cite{Baltz2}
\begin{equation}
\label{modpot}
\tilde{V_0}(x)=\tilde{V_3}(x)
\sim\delta (z-t)\left(\left(\frac{1}{x_\perp}\right)^{2iZ\alpha}-1\right)
\end{equation}
An inspection of (\ref{Ttot}) reveals that the scattering amplitude is
represented by a divergent integral. There are infrared divergencies caused
by the poles of the momentum transfer functions ${\cal T}_A$ and ${\cal
T}_B$, see Appendix \ref{approp}.
It is interesting to note, that an 
explicit introduction of a photon mass 
describing a screened Coulomb potential does {\it not} yield a regularized
expression for the functions ${\cal T}$. On the other hand, if 
the modified potential (\ref{modpot}) is screened with a damping factor
$e^{-\epsilon x_\perp}$, this  
leads in momentum space to 
\begin{equation}
\label{tscreened}
\tilde{V}(k)\sim
\frac{1}{(\epsilon^2+k_\perp^2)^{(1-i\alpha Z)}}\Gamma(2(1-i\alpha
Z))P_{1-2i\alpha Z}(\epsilon(\epsilon^2+ k_\perp^2)^{-\frac{1}{2}})
\end{equation}
which resembles the propagator of a photon with mass $\epsilon$.
$P_{1-2i\alpha Z}(\;\;)$ denotes a Legendre function. 

We should stress however, that such 
artificial regularization procedures are not needed. There is a natural cut
off since the
condition for the applicability of the used approximations requires (see
Appendix \ref{apotlim})
\[
\gamma \gg \frac{x_\perp}{|z-\beta t|}
\]
which in momentum space translates into the condition 
\[
k_\perp \gg \frac{\omega}{\gamma}
\]
This lower bound for the transverse momentum corresponds to the cut off
inherent in the Fourier transform of the potential, cf eq. (\ref{ftpot})
The introduction of this cut off has the important property to restore 
the energy dependence of the amplitude which was lost when taking the limit
$\gamma \to \infty$.

\subsection{Solution in the field of channeled ions}
\label{crystal}
Here we want to sketch briefly an extension of the formalism discussed so
far to the case of more that two colliding charges. 
The causal decoupling of interactions with sources moving on the positive
and negative light-cone, respectively, and the above interpretation of the
interaction 
can be used to calculate the
scattering amplitude of electrons (or more realistically electron-positron
pair production) for a field configuration which corresponds to the
channeling of an ion in a crystal. 

We use the equal speed
system, the crystal is moving in $-z$ direction.
The crystal
layers have a spatial distance $a\vec{e}_z$.
In the
ultrarelativistic case, the electron again interacts with the ion and the
crystal layers separately and we get simple time orderings of the
interaction. For $n$ crystal layers we have $n+1$ possibilities. 
For the sake of simplicity we formulate the perturbative description of the
successive interactions of the electron with both the ion and the crystal
layers directly with modified potentials of the form (\ref{modpot}). 
One then obtains for the interaction with two neighbouring crystal layers 
the integral
\begin{eqnarray}
A&=&-\delta (p'_+-p_+)e^{ip'_-\frac{a}{2}}\nonumber \\
&&\int\frac{dk_-d^2k_\perp}{(2\pi)^2}
\frac{\overline{u}(p')\hat{\gamma}_+u(p)}{-ik_-+
i\frac{k_\perp^2+m^2}{k_+}+\frac{\epsilon}{k_+}}e^{-ik_-\frac{a}{2}}{\cal
T}_{C_i}(\vec{k}_\perp-\vec{p}_\perp){\cal
T}_{C_{i+1}}(\vec{p'}_\perp-\vec{k}_\perp)\nonumber \\
&=&2\pi \delta (p'_+-p_+)
\int\frac{d^2k_\perp}{(2\pi)^2}e^{i\left(-\frac{k_\perp^2+m^2}{2p'_+}+
\frac{p'_-}{2}\right)a}\nonumber \\
&&{\cal T}_{C_i}(\vec{k}_\perp-\vec{p}_\perp)
{\cal T}_{C_{i+1}}(\vec{p'}_\perp-\vec{k}_\perp)
\overline{u}(p')\hat{\gamma}_+u(p)
\;\;\;\;a>0,\;\epsilon \to 0
\end{eqnarray}
The subscripts $C_i$ and $C_{i+1}$ denote the scattering amplitudes from the
interaction of the electron with the $i$th and the $(i+1)$th crystal layer.
For $a<0$ (reverse direction of electron motion) 
the integral vanishes, which expresses, that the electron
can not interact alternately with neighbouring crystal layers, due to
causality. The derivation of this functional connection using
(\ref{mtrafomi}) is shown in Appendix \ref{asolu2ion}.

If the electron interacts with the ion between interacting with two distinct
crystal layers, we get
\begin{eqnarray}
A&=& \int \frac{d^2k_\perp d^2k'_\perp}{(2\pi)^4}
 \frac{e^{ip'_-\frac{a}{2}}
\left(e^{-i ({k}_\perp^2
+m^2-i\epsilon )\frac{a}{2p_+}} -e^{-i ({k'}_\perp^2
+m^2-i\epsilon )\frac{a}{2p'_+}}\right)}{(p'_+({k}_\perp^2+m^2)-
p_+({k'}_\perp^2+m^2)-i \epsilon(p'_+-p_+))}
\nonumber \\
&&{\cal T}_{C_i}(\vec{k}_\perp-\vec{p}_\perp)
{\cal T}_A(\vec{k'}_\perp-\vec{k}_\perp)
{\cal T}_{C_{i+1}}(\vec{p'}_\perp-\vec{k'}_\perp)  \nonumber \\
&&\overline{u}(p')
(-\hat{\vec{\alpha}}_\perp \cdot \vec{k}_\perp +\hat{\gamma}_0m)
(-\hat{\vec{\alpha}}_\perp \cdot 
\vec{k'}_\perp +\hat{\gamma}_0m) \hat{\gamma}_+u(p)
\end{eqnarray} 
Successive
interactions with different crystal layers factorize and any scattering
process including intermediate interaction with the channeled ion gives the
same amplitude. Further studies will have to show how these considerations
can be put to use for the calculation of pair creation in channeling.

\section{Discussion}
\label{discussion}
In the previous sections the potential of a fast moving charge has been
substituted by its
asymptotic high-energy expression. 
From a mathematical (and also a physical) point of view this is a problematic
limit, since the required 
transformation is not an element 
of the Lorentz group. Furthermore, the potential 
(\ref{lwpot}), a bounded operator in Hilbert space, gets transformed 
into an unbounded operator, and finally the number of spatial dimensions gets
reduced from three to two.

The ansatz directly reflects the approximations made by Chang and Ma
\cite{Chang-Ma} 
who neglected the longitudinal components of the photon momentum, giving the
$\delta$-functions for the respective conserved light cone momenta. 
The above mentioned problems emerge here in the fact that the longitudinal
components of the photon momentum never really vanish.

All approximations allow the well known conclusion, that  
the eikonal expression can be regarded as 
the contribution of all ladder diagrams in the high-energy limit and that it 
is completely compatible with a perturbative calculation.

In the case of two ions we mainly profited from the causal decoupling of
the interactions implied by the presence of the factors $(1\pm \alpha_3)$. 
The matrix structure of the
true interaction is given by $(1\pm \beta\hat{\alpha}_3)\approx 
(1\pm \hat{\alpha}_3) \mp
\hat{\alpha}_3/2\gamma^2$, so that  
the leading corrections to this behaviour are suppressed with $1/\gamma^2$.

However, the considered calculations have  
inherent dangers. Fortunately 
we had to 
specify neither the transverse part of the potential nor its Fourier transform 
throughout our perturbative calculation. 
The first point may serve to generalize the validity of the result to any 
function $V_\perp(\vec{x}_\perp)$. However, a naive 
comparison even with the first order Born approximation 
would have failed due to the difficulties concerning the Fourier transform 
of the logarithm, 
whereas the logarithm in the argument of the exponential function 
is meaningful and correct. On the one hand 
the integration of the potential eliminates one dimension, 
which is finally recovered in the overall $\delta$ function for the light cone 
momenta. On the other hand 
the calculation of the scattering matrix between asymptotic states 
$(t\to \pm \infty)$ corresponds to the (unphysical) limit $\gamma \to \infty$. 
The detour via the Fourier transform of the ungauged potential gives a 
logarithm as well (see Appendix \ref{approp}), depending, however, 
strongly upon a regularization mass
$\mu$. This has its root in 
the fact, that $1/(k_\perp^2 + \mu^2)$ is not the 
correct two-dimensional photon propagator \cite{Grignani}. 
The divergent term $\lim_{\mu\to 0}\ln \mu^2$ is the term 
${\cal C}$ in section \ref{1iontrafo}. 

Now it is well known, that two-dimensional fields 
in the limit of vanishing mass 
are rather ill-defined objects, whereas the exponential of these fields is not. 
The 
Fourier transform of this exponential 
expression is elementary (see (\ref{smom})). 
It can further be expanded 
into a Taylor series. Although it is not justified to identify the different 
terms with the Fourier transforms of the powers of the logarithm, 
the first term corresponds to the high energy limit of the 
Fourier transform of the retarded potential, which is rather accidental. 

Nevertheless, the correct Fourier transform of the logarithm in two 
dimensions 
is obtained by Taylor expansion of ${\cal T}(k)$, 
but the limit $Z\alpha \to 0$ has to 
be taken {\it after} having integrated the expression with a test function
\cite{Grignani}. 
\begin{equation}
\label{lnft}
\int d^2x_\perp e^{-i\vec{k}_\perp\cdot \vec{x}_\perp}\ln x_\perp^2
=\lim_{iZ\alpha\to 0}\frac{d}{d(iZ\alpha)}\left(\pi \frac{\Gamma(1-i\alpha
Z)}{\Gamma(i\alpha
Z)}\left(\frac{4}{k_\perp^2}\right)^{1-i\alpha Z}
\right)
\end{equation}
This peculiarity is related to the fact, that the linearity of the Fourier
transform is not strictly defined for the action on infinite series,
resulting in the non-commutability of limiting procedures as in (\ref{lnft}). 
Another way is to rewrite the logarithm by the $t$ 
integral of the gauged potential and expressing the integrand by means 
of its Fourier transform. One then finds for the Fourier transform of the 
logarithm in two dimensions \cite{Ferrari}:
\begin{equation}
-\int d^2x_\perp e^{-i\vec{k}_\perp\cdot \vec{x}_\perp}
\ln x_\perp ^2 = \lim_{\lambda\to 0} 4 \pi \left( 
\frac{1}{k_\perp^2 +\lambda^2} + \pi \delta^2(k_\perp) \ln \left(\frac{
\lambda^2}{\mu^2}\right)\right)
\end{equation}
with $\lambda =\omega/\gamma$, $\mu = 2/e^C$. 
The condition $\lambda \to 0$ coincides with the limit $\gamma\to \infty$ 
(see Appendix \ref{approp}). 
The correct treatment of this result 
again requires the limit to be taken after integrating the expression with a 
test function. Inverse transformation shows the required 
independence of the result of the regularzation parameter $\lambda$. 

In view of the previous discussion one may infer that it is not justified to identify 
the eikonal expression using a gauged potential with first-order 
perturbation theory using the original potential, since this ignores the 
gauge transformation applied to the potential. 

In any case the above considerations show, that the limit $\gamma \to  \infty$ 
is pathological and that its implications should be studied with care. 

\section{Conclusion}
The transformation presented in section \ref{1iontrafo} directly yields the
scattering amplitude for (arbitrary) particles scattered at fast charge 
centers.
Due to Lorentz invariance the expression for the amplitude holds even for
the case of static scattering centers, which gives the classical form of the 
eikonal approximation. 
The essential ingredients in the static case -- the vanishing of the spin
current and the assumption that $\phi$ was a slowly varying function -- have
been replaced by the discontinuous behaviour at the light fronts in the
presented case of fast scattering centers. 

The gauge transformed high energy limit of the potential directly contains
the necessary approximations mentioned in section \ref{pertapp}, 
and perturbative calculations using this
potential can be done
without further assumptions. 

The results obtained by eikonal approximations or by application of the
transformation (\ref{mtrafomi}) are shown to be
equivalent to the sum of all ladder graphs. 
It should therefore not be surprizing that it is possible to regain 
perturbative results from
the eikonal expression.
We showed however, 
that the obtained results must be studied with care, and that a
wrong treatment only accidentally leads to correct results. 

Finally, the interaction of the electron with several ions moving along the
light cones was shown to decouple due to causality. 

The high-energy scattering amplitude of electrons in the field of two
colliding ions has the same structure as the second-order perturbative
result \cite{Bottcher}. It can therefore be considered as a two-photon
process with a modified
potential of the form (\ref{modpot}). The restrictions for the exchanged momenta imposed
by the considered approximations are compatible with those encountered in 
the Weizs\"acker--Williams method of virtual quanta. This
method corresponds to a first-order Born approximation in the temporal
gauge, considering only the transvere part of the interaction (the
longitudinal part is suppressed by $1/\gamma^2$). The analogy of the
obtained scattering amplitude to the
two-photon process allows to adopt the approximations made in
\cite{Bottcher}, aiming to express the result in the Weizs\"acker-Williams 
form. 
The cross section of the
scattering process can therefore be obtained from the Klein--Nishina formula
for Compton-scattering and photon distributions obtained from (\ref{modpot})
in the temporal gauge \cite{Eichmann}.
The simplified structure of the scattering amplitude allows for a study of
the high-energy behaviour of electron-positron pair production, that
accounts correctly for the Coulomb effects of both ions.

\section*{Acknowledgements}
U.E. would like to thank S.J.~Chang for helpful comments, and D.~Schwarz and
F.~Constantinescu for stimulating discussions. This work was supported by
{\it Deutsche Forschungsgemeinschaft} DFG within the project Gr-243/44-2.

\setcounter{equation}{0}
\renewcommand{\theequation}{\Alph{section}.\arabic{equation}}

\begin{appendix}
\section{Transformation of the Dirac equation}
\label{adiractrans}
According to (\ref{kovabl}) we set
\begin{eqnarray}
A_0&=& i\partial_t \ln \phi\\
A_3&=& i\partial_z \ln \phi'
\end{eqnarray}
Since $A_3=\beta A_0$ and $\partial_t =-\beta \partial_z$ due to the fact,
that the 
$t$ and $z$ dependence enters only via the combination ($z-\beta t$), we
find 
\begin{eqnarray}
&&\hphantom{\Rightarrow\;\;}\frac{\partial_z\phi'}{\phi'}=-\beta^2 
\frac{\partial_z\phi}{\phi}\\
&&\Rightarrow\;\; \phi'=\phi^{-\beta^2}
\end{eqnarray}
i.e. $\phi=1/\phi'$ for $\beta \to 1$ as expected (see
(\ref{kovabl}),(\ref{a0a3})).
Inserting this into the Dirac equation, we find 
\begin{eqnarray}
&&\left[ \phi \hat{\gamma}_0i\partial_t \frac{1}{\phi} + \phi
\phi^{-\frac{1}{\gamma^2}}\hat{\gamma}_3i\partial_z
\frac{1}{\phi\phi^{-\frac{1}{\gamma^2}}} +\hat{\vec{\gamma}}_\perp
i\vec{\nabla}_\perp -m\right]\psi\\
&&=\left[\phi\left( \hat{\gamma}_0i\partial_t + \phi^{-\frac{1}{\gamma^2}}
\hat{\gamma}_3
i\partial_z \frac{1}{\phi^{-\frac{1}{\gamma^2}}}\right)\frac{1}{\phi}
+\hat{\vec{\gamma}}_\perp
i\vec{\nabla}_\perp -m\right]\psi\\
&&=\phi\left[ \hat{\gamma}_0i\partial_t + \hat{\gamma}_3\left( 
i\partial_z -\frac{1}{\beta^2
\gamma^2}A_3\right) + \hat{\vec{\gamma}}_\perp (i \vec{\nabla}_\perp + i {\rm
grad}_\perp \ln \phi )-m \right]\tilde{\psi}\\
&&=0
\end{eqnarray}
In the last step we introduced $\tilde{\psi}=\psi/\phi$. 

In the ultrarelativistic limit terms of the order $1/\gamma^2$ are neglected. 
We end up with a Dirac equation coupled to a purely transverse
vector potential exhibiting a Heaviside step function dependence $\sim \theta
(t-z)$. These
properties are essential for our considerations. 

\setcounter{equation}{0}

\section{Ultrarelativistic limit of the potential}

\label{apotlim}
In this section we want to discuss the limit $\beta \to 1$ of the potential
(\ref{lwpot}).
From section (\ref{1iontrafo}) we expect the asymptotic form of the 
potential to 
be
\begin{equation}
\lim_{\beta \to 1}\frac{-1}{\sqrt{(z-t)^2+\frac{x_\perp^2}{\gamma^2}}} =
\delta (z-t) \ln (x_\perp^2) + {\cal C}'
\end{equation}
$t$ integration of (\ref{lwpot}) determines ${\cal C}'$
\begin{equation}
\label{cprime}
{\cal C}'=\frac{-1}{|z-t|}- \delta (z-t)\ln (\gamma^2)
\end{equation}
where we had to require $\gamma \gg x_\perp/|z-\beta t|$. \\
An attempt to derive the limit 
by means of Fourier transformation of (\ref{lwpot}) with respect to $z$ was
presented in \cite{Jackiw}. The Fourier transform reads 
\begin{eqnarray}
\int dz e^{i\omega z} \frac{1}{\sqrt{(z-\beta t)^2+\frac{x_\perp^2}{\gamma^2}}}
&=& e^{i\omega \beta t}\int dz e^{i\omega z}
\frac{1}{\sqrt{z^2+\frac{x_\perp^2}{\gamma^2}}}\nonumber \\
&=& 2e^{i\omega \beta t} K_0\left(\frac{\omega
x_\perp}{\gamma}\right)\nonumber \\
\label{FTlim}
&\stackrel{\gamma\to
\infty}{\longrightarrow}& -2e^{i\omega t}\ln
\left(\frac{\omega x_\perp e^C}{2 \gamma}\right)
\end{eqnarray}
The quantity $C$ here denotes Euler's constant.
The inverse Fourier transformation of this expression yields
\begin{equation}
\label{arbilim}
\lim_{\beta \to 1}\frac{1}{\sqrt{(z-t)^2+\frac{x_\perp^2}{\gamma^2}}}
=\frac{1}{|z-t|} + g(x_\perp)\delta (z-t)
\end{equation}  
The coefficient $g(x_\perp)$ of the delta distribution in this result is not
uniquely specified. Naive application of the textbook formula
\cite{Lighthill}
\begin{equation}
\label{lighthillf}
\int \frac{dk}{2\pi} \ln |k| e^{ikx} = -\frac{1}{2|x|}
\end{equation}
would give $g(x_\perp)=2[\ln(x_\perp/2\gamma)+C]$, but (\ref{lighthillf}) 
is valid only up to arbitrary multiples of $\delta (x)$.
The validity of (\ref{FTlim}), however, as well demands the 
condition $\gamma \gg \omega x_\perp$. 

It is possible to find a 
gauge transformation that removes both, 
the long-range potential 
$1/|z-t|$ as well as $\delta (z-t)\ln (\gamma^2)$ (\ref{cprime}). 
This is fulfilled with the gauge
transformation \cite{Aichelburg}
\begin{equation}
\label{gaugeob2}
\psi'=e^{iZ\alpha \ln\left(\gamma (z-t) +
\sqrt{1+\gamma^2(z-t)^2}\right)}\psi
\end{equation}
The gauge-transformed potential reads
\begin{equation}
\label{gpot}
A_0'=-\frac{Z\alpha\gamma}{\sqrt{\gamma^2(z-\beta t)^2 + x_\perp^2}} +
\frac{Z\alpha\gamma}{\sqrt{\gamma^2(z-\beta t)^2 + 1}}
\end{equation}
and has the ultrarelativistic limit 
\begin{equation}
\label{potlim}
\lim_{\beta\to 1}A_0'=+\delta (z-t) \ln (x_\perp ^2)
\end{equation}

The appearance of the logarithm follows immediately from the inhomogeneous
Maxwell equations in the Lorentz gauge that reduces to a two dimensional
Poisson equation in the limit $\beta\to 1$. 

This gauge transformation has the advantage to yield a short-range
potential that allows for asymptotic plane wave solutions. 
For this reason it was used to obtain a faster convergence in 
coupled channel calculations \cite{Eichler}.

\setcounter{equation}{0}
\section{Solution of the Dirac equation with two discontinuities}
\label{asolu2ion}
The Dirac equation for an electron moving in the field of two
ultrarelativistic colliding ions $A$ and $B$ reads
\begin{equation}
\label{de2ion}
\left[\frac{1}{2}\left(\hat{\gamma}_-i\partial_{\tau_-}+
\hat{\gamma}_+i\partial_{\tau_+}\right) +i\hat{\vec{\gamma}}_\perp
\cdot \vec{\nabla}_\perp -m -\frac{1}{2}\hat{\gamma}_-\delta (\tau_-) 
V^A_\perp -\frac{1}{2}\hat{\gamma}_+\delta (\tau_+) V^B_\perp \right] \psi=0
\end{equation}
where we used light cone variables $\tau_\pm=(t\pm z)/2$. 
One directly finds that 
$\hat{\gamma}_- \psi$ is discontinuous at $\tau_-$ through the action of ion
$A$ and $\hat{\gamma}_+\psi$ is discontinuous at $\tau_+=0$  through the
action of ion $B$, respectively. 

We introduce $\psi_\pm=(1\pm \hat{\alpha}_z)\psi$ and use
$2\psi=\psi_-+\psi_+$ to formulate the problem as follows
\begin{equation}
\left(i\partial_{\tau_+} +i\hat{\vec{\alpha}}_\perp\cdot \vec{\nabla}_\perp
-\gamma_0m  -\delta (\tau_+) V^B_\perp
\right)\psi_+ +\left(i\partial_{\tau_-}
+i\hat{\vec{\alpha}}_\perp\cdot \vec{\nabla}_\perp
-\gamma_0m -\delta (\tau_-) V^A_\perp\right)\psi_-=0
\end{equation}
where (\ref{de2ion}) has been multiplied by $2\hat{\gamma}_0$.
This we rewrite as
\begin{eqnarray}
&&\left(i\partial_{\tau_+} -\delta (\tau_+) V^B_\perp
\right)\psi_+ +\left(i\hat{\vec{\alpha}}_\perp\cdot \vec{\nabla}_\perp
-\gamma_0m\right)\psi_-\nonumber \\
\label{delambda}
&&=-\left(i\partial_{\tau_-}-\delta (\tau_-)
V^A_\perp\right)\psi_- -\left(i\hat{\vec{\alpha}}_\perp\cdot 
\vec{\nabla}_\perp
-\gamma_0m\right)\psi_+
\end{eqnarray}
By using the standard representation of Dirac matrices and 
simply rearranging the four equations (\ref{delambda}) one obtains
\begin{eqnarray}
\nonumber
&&\left[\left(\begin{array}{rr}1\!{\rm l}&0\\0&0\end{array}\right)
\left(i\partial_{\tau_+}
-\delta (\tau_+) V^B_\perp
\right)
+\left(\begin{array}{rr}0&0\\0&1\!{\rm l}\end{array}\right)
\left(i\partial_{\tau_-}-\delta (\tau_-)   
V^A_\perp\right)-
m\left(\begin{array}{rr}0&1\!{\rm l}\\1\!{\rm l}&0\end{array}\right)\right.\\
&&\left.+i\partial_x
\left(\begin{array}{rr}0&-\sigma_y\\ \sigma_y&0\end{array}\right)
+i\partial_y\left(\begin{array}{rr}0&-i\sigma_x\\i\sigma_x&0\end{array}\right)
\right]\tilde{\psi}=0\\
\nonumber
{\rm where}
&&\tilde{\psi}=\left(\begin{array}{c}\psi_1+\psi_3\\ \psi_2-\psi_4\\
\psi_1-\psi_3\\ \psi_2+\psi_4\end{array}\right)
\end{eqnarray}
corresponding to an isomorphic linear transformation \cite{Segev} with the
matrix 
\begin{equation}
\label{segevtrafo}
\Lambda=\left(\begin{array}{rr}
1\!{\rm l}&\sigma_z\\1\!{\rm l}&-\sigma_z\end{array}\right)
\end{equation}
Since $\Lambda$ is a bijection, 
each side of (\ref{delambda}) has to be zero. Off
the light fronts we therefore have the two equations
\begin{eqnarray}
\label{psi+}
i\partial_{\tau_+}\psi_+&=&(i\hat{\vec{\alpha}}_\perp\cdot \vec{\nabla}_\perp
-\gamma_0m)\psi_-\\
i\partial_{\tau_-}\psi_-&=&(i\hat{\vec{\alpha}}_\perp\cdot \vec{\nabla}_\perp
-\gamma_0m)\psi_+
\end{eqnarray}
According to (\ref{psiout}) the discontinuities at the light fronts are
described by 
\[
\psi_-(\tau_-=0^+)=\phi^A(x_\perp)\psi_-(\tau_-=0^-)\;\;,\;\;\;
\psi_-(\tau_+=0^+)=\phi^B(x_\perp)\psi_+(\tau_-=0^-)
\] 
$\phi^A$ and $\phi^B$ are defined by (\ref{phidef}) using the scalar parts of
the potentials of the ions $A$ and $B$.
Let us study the spinor $\psi_+$, evaluated at the surface $\tau_+=0^+$:
\begin{equation}
\label{taupgr0}
\psi_+(\tau_+=0^+)=\phi^B \frac{i\hat{\vec{\alpha}}_\perp\cdot 
\vec{\nabla}_\perp
-\gamma_0m}{p_-}\psi_-(\tau_+=0^-)
\end{equation}
In the region $\tau_->0$ 
the electron already has interacted with ion $A$ and we can write 
\begin{equation}
\label{taupmgr0}
\psi_+(\tau_+=0^+,\tau_->0)
=\phi^B \frac{i\hat{\vec{\alpha}}_\perp\cdot 
\vec{\nabla}_\perp
-\gamma_0m}{p_-}\phi^A (1-\alpha_z)\psi_p
\end{equation}
where $\psi_p$ is the incoming plane wave at momentum $p$.
This relation also can be obtained imeediately 
from (\ref{mtrafomi}) and (\ref{psi+})
for $\tau_+$ and $\tau_->0$. 
The operator $i\partial_+$ in (\ref{psi+}) has been replaced by its
eigenvalue $p_-$, the incoming negative light cone momentum. This is
possible since $p_-$ is 
conserved in the
interaction with ion $A$. 
The expansion of $\psi_+(\tau_+=0^+)$ in the plane-waves basis reads
\begin{equation}
\psi_+(\tau_+=0^+,\tau_->0)
=\int\frac{dp'_+d^2p'_\perp}{(2\pi)^3}B(p',p)e^{-ip'_+\tau_-
+i\vec{p'}_\perp\cdot \vec{x}_\perp}u(p')
\end{equation}
where we substituted $d^3p' \to dp'_+d^2p'_\perp$ \cite{Meggiolaro}.
According to (\ref{taupgr0}) the expansion coefficients are 
\begin{equation}
B(p',p)=\int_0^\infty d\tau_-\int d^2x_\perp e^{ip'_+\tau_-
-i\vec{p'}_\perp\cdot \vec{x}_\perp}\phi^B
\overline{u}(p')\frac{i\hat{\vec{\alpha}}_\perp\cdot \vec{\nabla}_\perp
-\gamma_0m}{p_-}\psi_-(\tau_+=0^-,\tau_->0)
\end{equation}
In the region $\tau_->0$, $\tau_+<0$ the wave function $\psi_-$ is a freely
propagating wave packet with a fixed light cone momentum $p_-$ and a
superposition of transverse momenta $p_\perp$. The mass shell condition
requires $p_-p_+=\vec{q}^2_\perp +m^2$. In this way
$\psi_-(\tau_+<0,\tau_->0)$ can be obtained from
$\psi_-(\tau_+<0,\tau_-=0^+)$. We have 
\begin{equation}
\psi_-(\tau_->0)=
\int\frac{d^2q_\perp}{(2\pi)^2}
e^{-i\frac{q_\perp^2+m^2}{p_-}\tau_--ip_-\tau_++
i\vec{q}_\perp\cdot \vec{x}_\perp}
\int d^2x'_\perp e^{i\vec{x'}_\perp\cdot 
(\vec{p}_\perp-\vec{q}_\perp)}\phi^A
(1-\hat{\alpha}_z)u(p)
\end{equation}
which leads to 
\begin{eqnarray}
B(p',p)&=&i\int \frac{d^2q_\perp}{(2\pi)^2} \int d^2x'_\perp
e^{i\vec{x'}_\perp\cdot (\vec{q}_\perp-\vec{p'}_\perp)}\phi^B\int d^2x_\perp
e^{i\vec{x}_\perp\cdot (\vec{p}_\perp-\vec{q}_\perp)}\phi^A\nonumber \\
\label{bexp}
&&\overline{u}(p')
\frac{\hat{\vec{\alpha}}_\perp\vec{q}_\perp -\gamma_0m}
{p'_+p_--q_\perp^2-m^2+i\epsilon}(1-\hat{\alpha}_z)u(p)
\end{eqnarray}
Note, that the lower bound of the $\tau_-$ integration is 0, since we
inserted the expression of $\psi_-$ for $\tau_->0$. 

Together with the corresponding term for the reverse order of interactions
with the two ions, (\ref{bexp}) 
is the $S$ matrix for an electron scattered at the light
fronts, first derived by Segev and Wells \cite{Segev} in an elegant way using
the transformation (\ref{segevtrafo}).

If both ions $A$ and $B$ move on positive light cones separated by
the spatial distance $a\vec{e}_z$ (see section \ref{crystal}), we obtain with
(\ref{mtrafomi}) for the interacting part of the spinor $\psi$
\begin{eqnarray}
\psi_+(\tau_+-a/2=0^+)&=&\phi^A\psi_+(\tau_+-a/2=0^-)\nonumber \\
&=&\phi^A\int\frac{d^2q_\perp}{(2\pi)^2}
e^{-i\frac{q_\perp^2+m^2}{p_+}\frac{a}{2}-ip_+\tau_-
+i\vec{q}_\perp\cdot \vec{x}_\perp}\nonumber \\
&&\int d^2x'_\perp e^{i\vec{x'}_\perp\cdot 
(\vec{p}_\perp-\vec{q}_\perp)}\phi^B
(1+\hat{\alpha}_z)u(p)
\end{eqnarray}
The expansion of $\psi_+$ in plane waves at the point
$\tau_+=a/2+0^+$
yields the $S$ matrix of this process in momentum space
\begin{eqnarray}
S(p',p)&=&2\pi \delta(p'_+-p_+)i\int \frac{d^2q_\perp}{(2\pi)^2} 
e^{i\left(-\frac{q_\perp^2+m^2}{2p'_+}+\frac{p'_-}{2}\right)a}\nonumber \\
&& \int d^2x_\perp e^{i\vec{x}_\perp\cdot (\vec{q}_\perp-\vec{p'}_\perp)}
\phi^A \int d^2x_\perp
e^{i\vec{x}_\perp\cdot (\vec{p}_\perp-\vec{q}_\perp)}\phi^B \overline{u}(p')
(1+\hat{\alpha}_z)u(p)
\end{eqnarray}
in accordance with section \ref{crystal}.

\setcounter{equation}{0}
\section{The photon propagator at high collision energies}

\label{approp}
The four-dimensional 
Fourier transform of the potential (\ref{lwpot}) reads
\begin{equation}
\label{ftpot}
\int d^4x e^{ikx}\frac{-Z\alpha \gamma}{\sqrt{\gamma^2(z-\beta t)^2 +
\vec{x}^2_\perp}}=-(2\pi)^2 Z \alpha \delta (k_0-\beta k_3)
\frac{2}{\left(\frac{k_3}{\gamma}\right)^2 + k_\perp^2}
\end{equation}
which has the following low and high-velocity limits 
\begin{eqnarray}
\lim_{\beta \to 0}&=&-(2\pi)^2 Z \alpha \delta (k_0)\frac{2}{|\vec{k}|^2}\\
\label{helimft}
\lim_{\beta \to 1}&=&-(2\pi)^2 Z \alpha \delta (k_0-k_3)\frac{2}{k^2_\perp}
\end{eqnarray}

The last expression reflects the observation, that in the high-energy limit
the longitudinal components $k_-$ and $k_+$ of the photon momentum can be
dropped.

After having performed the gauge transformation (\ref{gaugeob2}) and taken
the limit $\gamma \to \infty$, the
potential to be transformed is expression (\ref{potlim}). 
Grignani and Mintchev \cite{Grignani} have shown, that it is wrong to
identify the Fourier transform of (\ref{potlim}) with (\ref{helimft}) or
with the
regulated expression $1/(k_\perp^2 + \mu^2)$ with a regulating mass inserted
by hand. 

Calculating the time integral of $A_0$ in the eikonal expression
and using (\ref{ftpot}) one finds
\begin{equation}
\int^\infty_{-\infty}dt A_0 =Z\alpha \ln (x_\perp^2) +Z\alpha 
\lim_{\mu\to 0} \ln \mu^2
\end{equation}
with $1/2e^C$ absorbed in $\mu$ as in \cite{Jackiw}. The term 
$\lim_{\mu\to 0} Z\alpha\ln \mu^2$ is
the term ${\cal C}$ in section \ref{1iontrafo} and is 
completely different from $\mu$ in eq (7) of \cite{Jackiw}! 

One may attempt to calculate the two-dimensional Fourier transform of the
logarithm 
from a Taylor expansion in powers of $iZ\alpha$ of the Fourier transform of
the $T$ matrix (i.e. its transverse part) which is given by the following
closed expression 
\begin{eqnarray}
{\cal T}(k_\perp)&=&
\left(\frac{4}{k^2_\perp}\right)^{1-i\alpha Z}\Gamma^2(1-i\alpha Z)
\sin(\pi i\alpha Z)\\
&=&\pi \frac{\Gamma(1-i\alpha Z)}{\Gamma(i\alpha
Z)}\left(\frac{4}{k^2_\perp}\right)^{1-i\alpha Z}  
\end{eqnarray}
The first terms of the Taylor expansion read 
\begin{eqnarray}
{\cal T}(k_\perp)&\approx& +4\pi i\alpha Z
\frac{1}{k^2_\perp} + 4\pi (i\alpha Z)^2 \frac{\ln
(k^2_\perp/4)+C}{k^2_\perp}\nonumber \\
\label{ttaylor}
&&+ 2\pi
(i\alpha Z)^3 \frac{\ln^2(k^2_\perp/4)+ 4C\ln(k^2_\perp/4)+4C^2}
{k^2_\perp} + \dots
\end{eqnarray}
The second term would then correspond to the desired Fourier transform
(times $(iZ\alpha)$),
the third term correspondingly to $(iZ\alpha)^2$ times 
the Fourier transform of the square of the
transverse part of the potential (\ref{potlim}) that has to be compared
with the result of Torgerson \cite{Torgerson}. 

This is, however, not justified, since the linearity of the Fourier transform
is only guaranteed for finite sums and causes problems when applied to
infinite series like the Taylor expansion of the exponential function. 
To get the correct result for the exact two dimensional euclidean
photon propagator, the limit $iZ\alpha \to 0$ in 
\begin{equation}
\label{2dimpott}
\int d^2x_\perp e^{-i\vec{k}_\perp\vec{x}_\perp}\ln x_\perp^2
=\lim_{iZ\alpha\to 0}\frac{d}{d(iZ\alpha)}\left(\pi \frac{\Gamma(1-i\alpha
Z)}{\Gamma(i\alpha
Z)}\left(\frac{4}{k_\perp^2}\right)^{1-i\alpha Z}
\right)
\end{equation}
has to be taken after having
integrated the result with a test function. Performing the limit without
this precaution 
gives the wrong result (\ref{helimft}). 

Another form of the correct Fourier transform was derived in 
\cite{Ferrari}. We
obtain the equivalent form from the gauged potential $A'_0$ in (\ref{gpot}). 
Since
\begin{equation}
\label{lnpotdef}
-\ln x_\perp^2=\lim_{\gamma\to \infty}\int_{-\epsilon}^\epsilon dt \left(
\frac{\gamma}{\sqrt{\gamma^2t^2+x_\perp^2}}-
\frac{\gamma}{\sqrt{\gamma^2t^2+1}} \right)
\end{equation}
($\epsilon$ is arbitrary but finite)
and 
\begin{equation}
\int dtd^2x_\perp e^{i\omega t-i\vec{k}_\perp\cdot\vec{x}_\perp} \left(
\frac{\gamma}{\sqrt{\gamma^2t^2+x_\perp^2}}-
\frac{\gamma}{\sqrt{\gamma^2t^2+1}} \right)=4\pi\left(
\frac{1}{\left(\frac{\omega}{\gamma}\right)^2+k_\perp^2}-2\pi
K_0\left(\frac{\omega}{\gamma}\right)\right)
\end{equation}
we find by direct substitution
\begin{equation}
-\int d^2x_\perp e^{-i\vec{k}_\perp\vec{x}_\perp}\ln x_\perp^2
=\lim_{\lambda \to 0}4\pi\left(\frac{1}{k_\perp^2+ \lambda^2}+\pi\delta^2
(k_\perp)\ln\left(\frac{\lambda^2}{\mu^2}\right)\right)
\end{equation}
with $\lambda =
\omega/\gamma$, $\mu=2/e^C$. 
The limit has to be treated in the same way as in (\ref{2dimpott}).

When naively taking the limit $iZ\alpha \to 0$ immediately, by chance 
one obtains the high energy limit of the ungauged potential. 

\end{appendix}

\end{document}